\PassOptionsToPackage{table,xcdraw}{xcolor}
\documentclass{IEEEmce}

\usepackage[colorlinks,urlcolor=blue,linkcolor=blue,citecolor=blue]{hyperref}

\usepackage{upmath}
\usepackage{graphicx}
\usepackage{subfigure}
\usepackage{capt-of,etoolbox}
\usepackage{svg}
\usepackage{dblfloatfix}

\usepackage{amsmath}
\usepackage{amsfonts}
\usepackage{graphicx}
\usepackage{subfigure}
\usepackage{multirow}
\usepackage{multicol}
\usepackage{lipsum}
\usepackage{cite}
\usepackage{amssymb}
\usepackage{balance}
\usepackage{tabularx,ragged2e,booktabs}
\usepackage{multicol,tabularx,capt-of}
\usepackage{multirow}
\usepackage{lettrine}
\usepackage{tikz}
\usepackage[export]{adjustbox}

\usepackage[T1]{fontenc}
\usepackage{multirow}

\usepackage[T1]{fontenc}

\usepackage[table,xcdraw]{xcolor}

\jvol{XX}
\jnum{XX}
\paper{8}
\jmonth{May/June}
\publisheddate{00 xxxx 0000}
\currentdate{00 xxxx 0000}
\jname{IEEE Consumer Electronics Magazine}
\pubyear{2022}
\doiinfo{MCE.2022.Doi Number}

\begin{document}

\sptitle{Department: Head}
\editor{Editor: Name, xxxx@email}

\title{CAN Bus: The Future of Additive Manufacturing (3D Printing)}

\author{Jun-Cheng Chin}
\affil{University of Tennessee, Knoxville, US}

\author{Himanshu Thapliyal}
\affil{University of Tennessee, Knoxville, US}

\author{Tyler Cultice}
\affil{University of Tennessee, Knoxville, US}

\markboth{Department Head}{CAN Bus: The Future of Additive Manufacturing (3D Printing)}

\begin{abstract}
Additive Manufacturing (AM) is gaining renewed popularity and attention due to low-cost fabrication systems proliferating the market. Current communication protocols used in AM limit the connection flexibility between the control board and peripherals; they are often complex in their wiring and thus restrict their avenue of expansion. Thus, the Controller Area Network (CAN) bus is an attractive pathway for inter-hardware connections due to its innate quality. However, the combination of CAN and AM is not well explored and documented in existing literature. This article aims to provide examples of CAN bus applications in AM.
\end{abstract}

\maketitle

\enlargethispage{10pt}

\chapterinitial{Additive Manufacturing's} (AM) place in the manufacturing industry is indisputable with the advent of low-cost solutions for parts fabrication, often known to the masses under the alias "3D Printing". The cost of 3D printing has dramatically decreased, leading to the advancement and widespread adoption of this technology. According to \cite{AA1}, approximately 2.2 million 3D printers were sold globally in 2021, with the market value expected to increase by 20.8\% between 2022 and 2030. The RepRap movement brought about a revolution in AM with its mission of developing machines that clone themselves with self-fabricated parts, cheap electronics, and servos \cite{BB1}. As the name would suggest, AM is a technique where the material is deposited in layers to form a 3-D structure. The most popular type of 3D printer among enthusiasts and professionals is one that utilizes the material extrusion process. Other AM processes include binder jetting, directed energy deposition, material jetting, powder bed fusion, sheet lamination, and vat photopolymerization. 
\begin{figure}[h]
\vspace*{-10pt}
	\centering \includegraphics[trim=1.95cm 5.3cm 1.9cm 6.6cm, scale=0.35, clip]{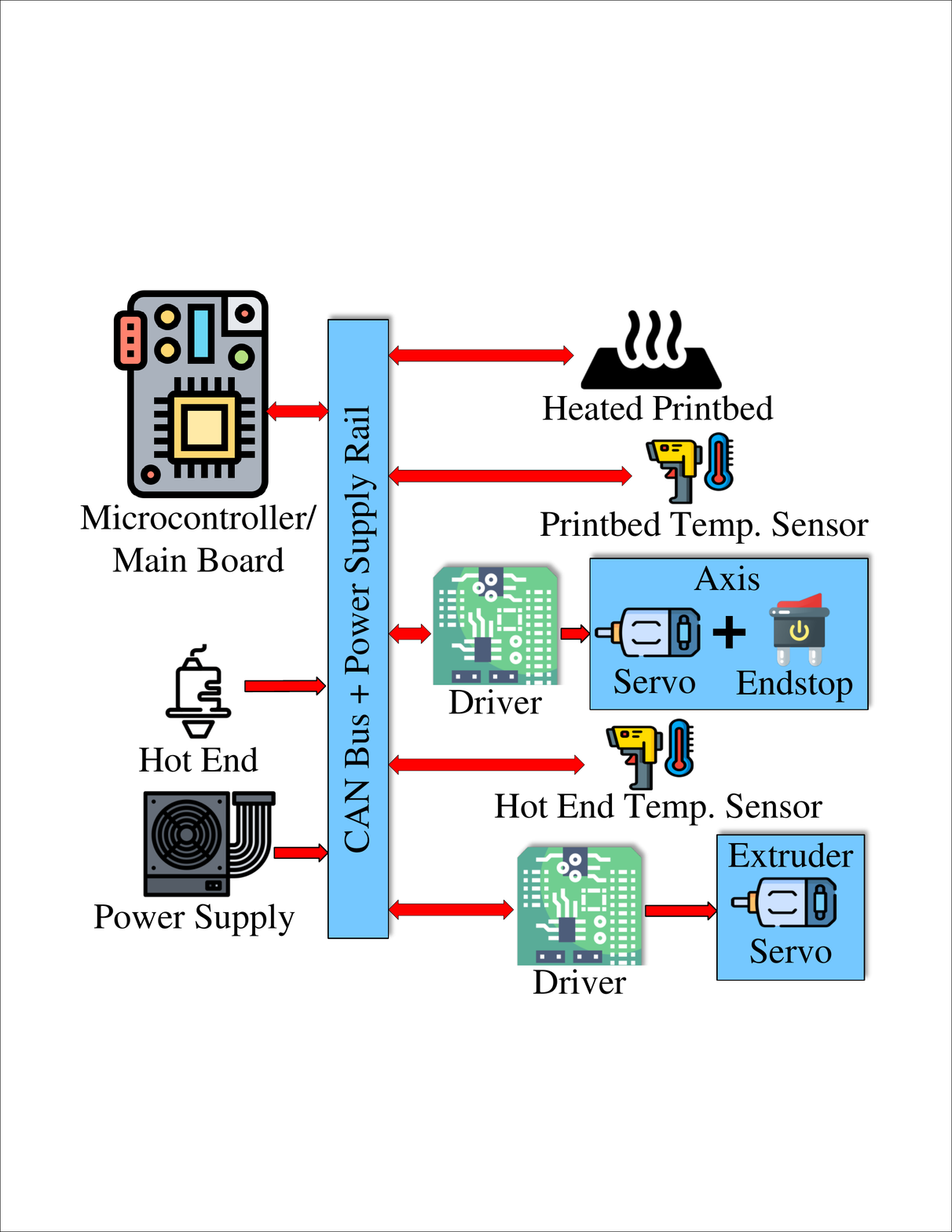}
	\caption{General idea of 3D Printer with CAN bus.}	
	\label{fig:scheme}
\end{figure}

The Controller Area Network (CAN) communication protocol is widely used in the automotive industry, dating back to the early 1990s. Nowadays, the majority of vehicles on the market implement CAN due to their robustness and reliability \cite{CC1}. Commercial vehicles may contain hundreds of Electrical Control Units (ECU), or controllers, which operate the vehicle. CAN enable the integration of all of these ECUs without incurring costs, speed, or integrity penalties. Adoption of the CAN bus has also seen massive interest from the aerospace industry \cite{DD1, EE1}. The goal is to replace or improve current communication protocols to provide this reduction in cost, reliability, and functionality in various embedded systems \cite{FF1}. All of these benefits can be applied to simplify and increase the utility of AM control systems. 

Popular AM machines in the current market, sought by enthusiasts and labs, use Fused Filament Fabrication (FFF) or Fused Deposition Modeling (FDM) type \cite{GG1}. The 3D printer's heart, or the microcontroller, is generally cost-effective and has low processing power in nature. The basic job of a 3D printer controller can be boiled down to servo or motor controls and temperature detection. Serial Peripheral Interface (SPI), Inter-Integrated Circuit (I2C), and Universal Asynchronous Receiver Transmitter (UART) are used for inter-hardware communication and control. Table 1 provides a general rundown of inter-hardware communications commonly used in 3D printers \cite{ZZ1}. These devices utilize point-to-point communication systems, requiring individual interconnections for each implemented device. The complexity of upgrading the system increases depending on its scale and functionality. Additionally, the various sensors and motor drivers connecting to the platform require fast and robust messages with low error tolerances.  

The attractiveness of CAN lies in its integrity-based design, which can achieve extremely fast data payloads. Fig. \ref{fig:scheme} gives an illustration of the general idea of a 3D printer with a CAN bus. In addition, it is a multi-master and multi-destination system all within a single bus, enabling avenues for modularity or plug-and-play capabilities. The conventional 3D printer has an exponential wire connection. With the CAN bus, the wire numbers will be reduced to a constant number \cite{YY1}. In effect, the CAN bus improves the ease of modularity and part-cost disparity.  

This article seeks to highlight and provide examples of CAN bus applications in AM for different scopes and scenarios. We have considered four case studies: (i) CAN bus integration within an Internet of Things (IoT) 3D printing farm, (ii) CAN connected immediate part adhesion testing with a test instrument add-on, (iii) mobile large-scale cement 3D printing with CAN, and (iv) novel fiber orientation control using a CAN connected rotational print head. The rest of this article is organized as follows: Section 2 goes through a brief background of CAN, laying out the wiring configuration, message structure, arbitration, and error correction mechanisms. Next, Section 3 highlights five examples of CAN bus applications in additive manufacturing. Finally, section 4 discusses future research directions of CAN in AM and concerns with the major adoption of the communication protocol.

\section{BACKGROUND ON CAN PROTOCOL}
\begin{figure}[h]
\vspace*{-8pt}
	\centering	\includegraphics[trim=1.2cm 4cm 1.2cm 3.2cm, width=3.05in, clip]{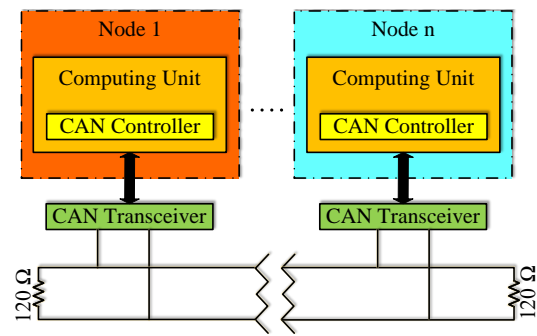}
	\caption{CAN wiring configuration \cite{HH1,II1,JJ1,KK1,LL1}.}	
	\label{fig:CANwiring}
\end{figure}

\begin{table*}[]
\centering
\begin{tabular}{|
>{\columncolor[HTML]{FFF2CC}}l 
>{\columncolor[HTML]{FFF2CC}}l 
>{\columncolor[HTML]{FFF2CC}}l 
>{\columncolor[HTML]{FFF2CC}}l 
>{\columncolor[HTML]{FFF2CC}}l 
>{\columncolor[HTML]{FFF2CC}}l |}
\hline
\multicolumn{6}{|l|}{\cellcolor[HTML]{C65911}{\color[HTML]{FFFFFF} \textbf{Table   1. Comparison between common communication protocols and CAN}}}                                                                                                                                                                                                                                                                                                                                                                                        \\ \hline
\multicolumn{1}{|l|}{\cellcolor[HTML]{FFD966}\textbf{Protocol}} & \multicolumn{1}{l|}{\cellcolor[HTML]{FFD966}\textbf{Complexity}}                                                                & \multicolumn{1}{l|}{\cellcolor[HTML]{FFD966}\textbf{Speed}}                                                       & \multicolumn{1}{l|}{\cellcolor[HTML]{FFD966}\textbf{No. of Wires}} & \multicolumn{1}{l|}{\cellcolor[HTML]{FFD966}\textbf{No. of Devices}}               & \cellcolor[HTML]{FFD966}\textbf{No. of master \& slave} \\ \hline
\multicolumn{1}{|l|}{\cellcolor[HTML]{FFF2CC}\textbf{UART}}     & \multicolumn{1}{l|}{\cellcolor[HTML]{FFF2CC}\textbf{Simple}}                                                                    & \multicolumn{1}{l|}{\cellcolor[HTML]{FFF2CC}\textbf{Slowest}}                                                     & \multicolumn{1}{l|}{\cellcolor[HTML]{FFF2CC}\textbf{1}}            & \multicolumn{1}{l|}{\cellcolor[HTML]{FFF2CC}\textbf{Up to 2}}                      & \textbf{Single to Single}                               \\ \hline
\multicolumn{1}{|l|}{\cellcolor[HTML]{FFF2CC}\textbf{I2C}}      & \multicolumn{1}{l|}{\cellcolor[HTML]{FFF2CC}\textbf{\begin{tabular}[c]{@{}l@{}}Easy to chain \\ multiple devices\end{tabular}}} & \multicolumn{1}{l|}{\cellcolor[HTML]{FFF2CC}\textbf{\begin{tabular}[c]{@{}l@{}}Faster \\ than UART\end{tabular}}} & \multicolumn{1}{l|}{\cellcolor[HTML]{FFF2CC}\textbf{2}}            & \multicolumn{1}{l|}{\cellcolor[HTML]{FFF2CC}\textbf{Up to 127 , but gets complex}} & \textbf{Multiple slaves \& master}                      \\ \hline
\multicolumn{1}{|l|}{\cellcolor[HTML]{FFF2CC}\textbf{SPI}}      & \multicolumn{1}{l|}{\cellcolor[HTML]{FFF2CC}\textbf{\begin{tabular}[c]{@{}l@{}}Complex as \\ device increases\end{tabular}}}    & \multicolumn{1}{l|}{\cellcolor[HTML]{FFF2CC}\textbf{Fast}}                                                        & \multicolumn{1}{l|}{\cellcolor[HTML]{FFF2CC}\textbf{4}}            & \multicolumn{1}{l|}{\cellcolor[HTML]{FFF2CC}\textbf{Many, but get complex}}        & \textbf{1 master, multiple slaves}                      \\ \hline
\multicolumn{1}{|l|}{\cellcolor[HTML]{FFF2CC}\textbf{CAN}}      & \multicolumn{1}{l|}{\cellcolor[HTML]{FFF2CC}\textbf{\begin{tabular}[c]{@{}l@{}}Easy to chain \\ multiple devices\end{tabular}}} & \multicolumn{1}{l|}{\cellcolor[HTML]{FFF2CC}\textbf{Fast}}                                                        & \multicolumn{1}{l|}{\cellcolor[HTML]{FFF2CC}\textbf{2}}            & \multicolumn{1}{l|}{\cellcolor[HTML]{FFF2CC}\textbf{Many, but simple}}             & \textbf{All of them are Masters}                        \\ \hline
\end{tabular}
\vspace*{-15pt}
\end{table*}

\begin{figure*}[b]
\vspace*{-10pt}
	\centering	\includegraphics[trim=0.5cm 6cm 0.5cm 5.8cm, scale=0.53, clip]{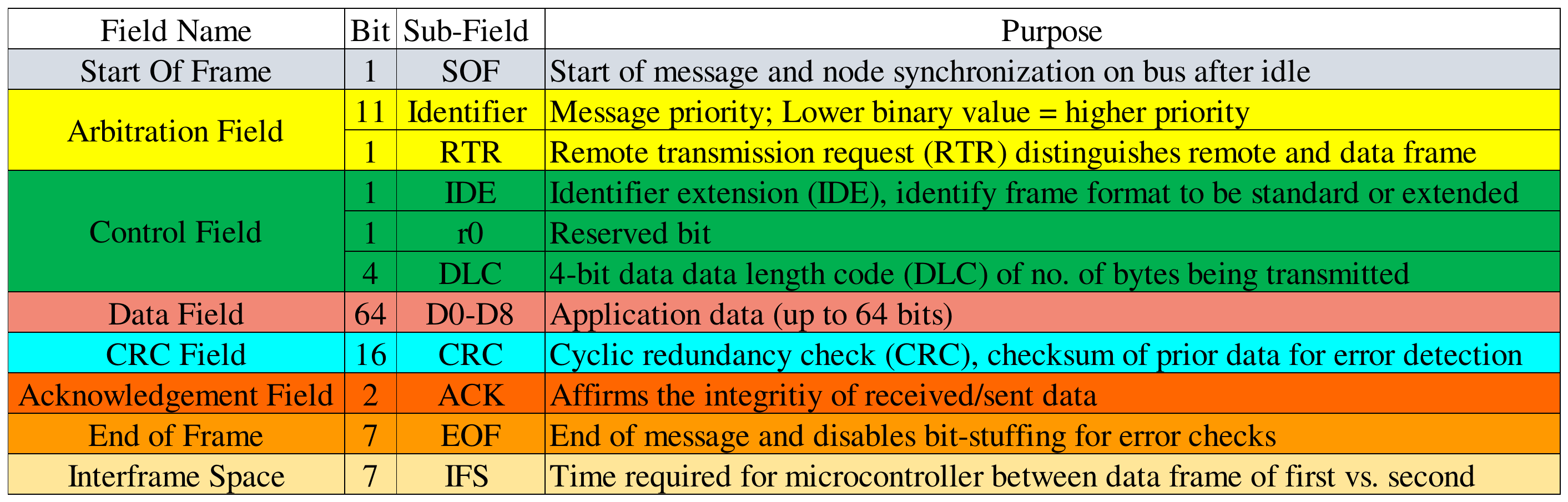}
		\caption{Summary of Standard CAN message frame bit field\cite{JJ1}.}	
	\label{fig:CAN_message}
\end{figure*}

The building blocks of CAN hardware design consist of a computing unit, a CAN controller, and a CAN transceiver. Each connected device containing all these components forms a CAN Node and communicates over a differential pair of bus wires. Even as the node count increases, the multiplexed/arbitrated bus design allows all CAN nodes to communicate along the same shared pair of wires, as shown in fig. \ref{fig:CANwiring}.  

Conventional multi-wire loom connections, which span up to thousands of wires, are reduced to a single bus with CAN. This also effectively reduces the price of wires and the weight of the product, resulting in an overall reduction in development costs. The two-wire bus uses a differential wiring mode (CAN-H and CAN-L) that reduces wiring complexity, noise, and electrical interference in noisy environments. Furthermore, the CAN bus's effective distance correlates with the data transfer rate. The CAN standard provides various maximum wire lengths, from 40 m at 1 Mbps to 1 km at 10 kbps. 
   
The standard's specification details the structure of the message frame and error/fault detection \cite{MM1}. CAN message packets or frames have four fields: arbitration, control, data, and cyclic redundancy check (CRC). These fields must be assigned properly for successful CAN message transmission. Fig. \ref{fig:CAN_message} summarizes the standard CAN message frame with its bit field assignment and purpose. 

Unlike master-slave protocols like the I2C interface or SPI, CAN is a multi-master serial communication protocol. This means every CAN node is allowed to communicate with each other, regardless of hierarchy. In addition, all messages are destination or message-ID based, as CAN arbitration will broadcast them throughout the whole bus regardless of the source of transmission. Thus, all nodes must actively 'listen' to messages on the bus but will only act when arbitration/ID criteria are met. While ID-based arbitration contains no source checking or secure end-to-end communication, this method is very attractive for data monitoring and multi-destination purposes. 

What makes CAN reliable rests in its built-in bit-wise arbitration and error checks. Under the hood, bit-wise arbitration allows the nodes to choose whether to ignore or accept the message based on the 11-bits of ID and the ID mask. Additionally, dominant IDs can prioritize messages when two nodes transmit data at the same time, with the unprioritized message attempting to re-transmit shortly after. This is great as lower priority nodes automatically retry transmission and do not waste resources. The error checking mechanisms for CAN are performed on the physical and digital level\cite{LL1}. Physical checks are handled with an error confinement mechanism (ECM) within every node, where faulty ones are disabled from the bus traffic upon detection. On the digital level, the CAN specification details five error types: bit, stuffing, CRC, form, and acknowledgement\cite{MM1}.   

\section{CASE STUDIES REGARDING APPLICATION OF CAN IN AM}
Advertisement of the general utilization of CAN for inter-hardware communication by commercial 3D printer manufacturers is rare. One example is the Snapmaker 2.0 by Snapmaker, a 3D printer that applies CAN for its 3-in-1 printing platform. Their machine is capable of Fused Filament Fabrication (FFF), CNC milling, and laser engraving. Depending on the application, the printer's printing heads, along with its respective print beds, are easily interchangeable. This model is a departure from the original that uses RJ45s and RJ25s, which requires a point-to-point connection and a dedicated connector for each component.  

Reviewing the available literature, the adoption of CAN in AM is an emerging topic and is very promising. The rest of this article sheds light on examples of applications of CAN by different groups in AM.   

\subsection{Case Study 1: 3D Printing Monitoring Platform Based On the IoT}
\begin{figure}[!ht]
\vspace*{-10pt}
	\centering	\includegraphics[trim=0.5cm 7.5cm 0.5cm 6.4cm, width=2.8in, clip]{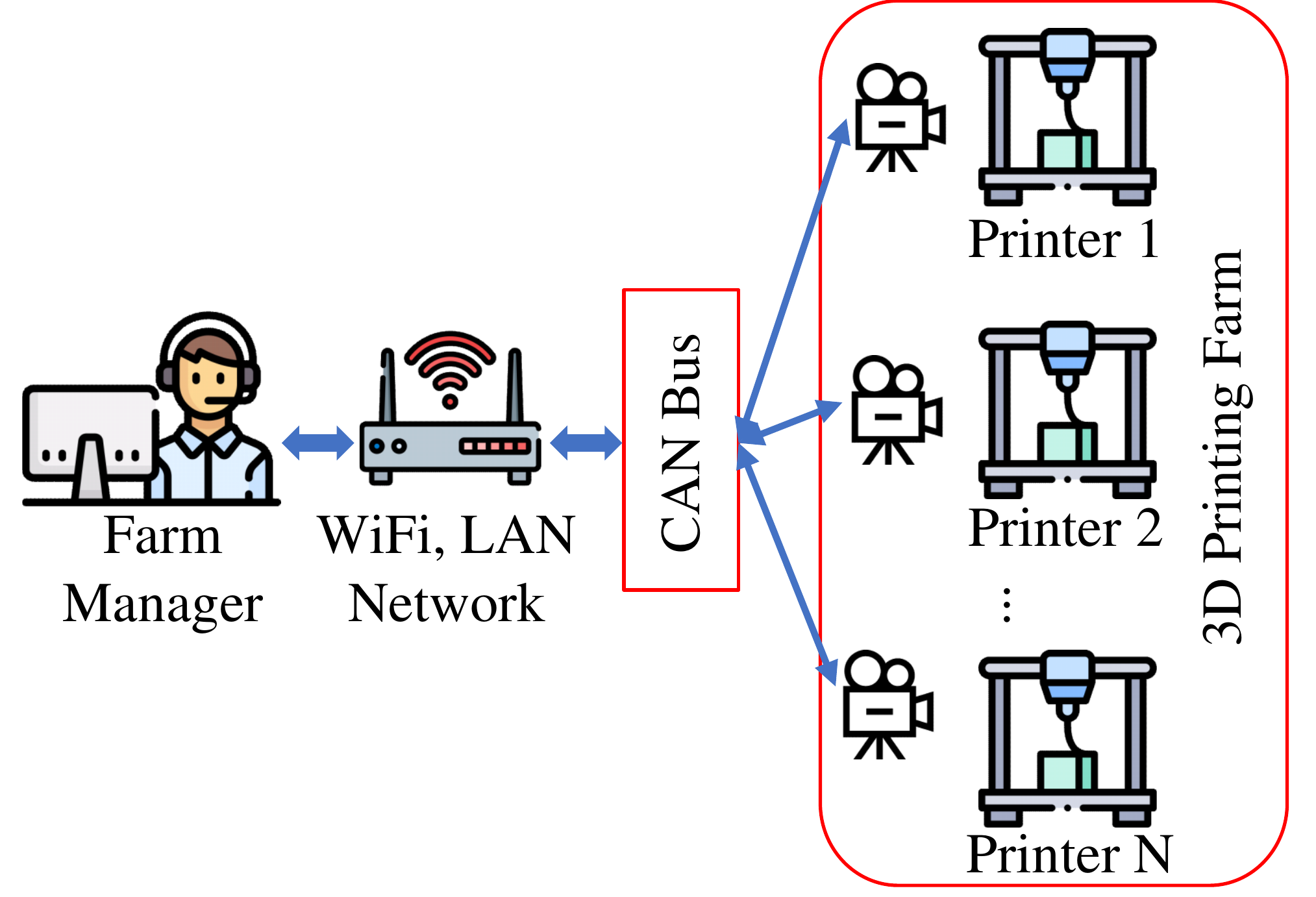}
	\caption{3D Printing Farm with CAN bus camera link \cite{NN1}.}	
	\label{fig:3D_IOT_FARM}
\vspace*{-10pt}
\end{figure}
A 3D printer farm is a large collection of printers designed with the objective of mass producing what a normal printer is limited to. Depending on the 3D model, material, and resolution, a normal print job could take roughly 10 minutes to more than 24 hours. It is imperative to monitor the status of the printer and printed parts. Monitoring is compounded and made extremely crucial as the system scales to the size of a farm. Every node of printers must be observed to ensure maximum profit and time investment.  
 

The main structure of a 3D printer farm features a central computer connected to a hub of 3D printers and an operator to monitor the operation. In J.J. Wu et al's work, integration of the CAN bus as part of the data monitoring node is demonstrated within a wireless network \cite{NN1}. Fig. \ref{fig:3D_IOT_FARM} shows an illustration of a 3D printing farm with the CAN bus camera link. The multi-master nature of the CAN protocol allows all nodes to broadcast all-in-one information within the bus, which fulfills the objective of quality control. In addition, the complexity of inner-printer connections is drastically reduced. However, the limitations of their printers caused only a partial CAN solution where the print quality is monitored via CAN-connected cameras.   

\subsection{Case Study 2: Part Adhesion Measuring Device in FFF process}
\begin{figure}[!ht]
\vspace*{-10pt}
	\centering	\includegraphics[trim=1cm 3.5cm 1cm 2cm, width=2.9in, clip]{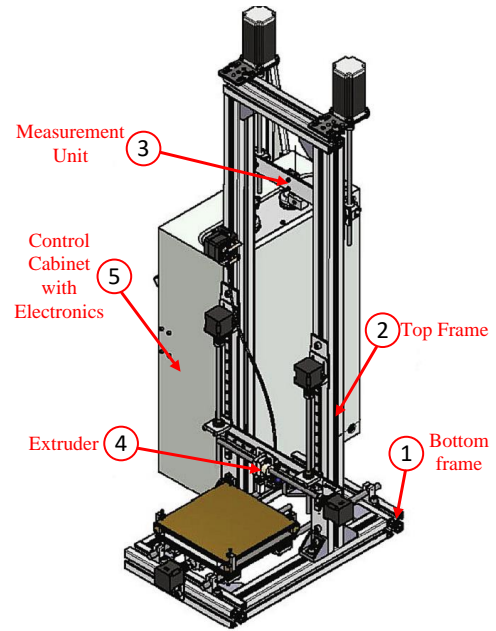}
	\caption{Model of FFF 3D Printer with Tensile Testing Machine by  \cite{OO1} is licensed under CC BY 4.0.}	
	\label{fig:tensile3d}
\end{figure}
The low cost and rapid prototyping of these printers are highly sought-after by laboratories around the world. Depending on the print material, adhesion to the bed surface may come into question. Despite the mass proliferation of these printers, there do not seem to be standards that provide guidelines for compatible materials for print bed adhesion. 

Laumann, Daniel, et al. constructed a 3D printing machine with an integrated tensile testing device \cite{OO1}. The main goal is to directly test the adhesion forces between the printed part and the print bed immediately after completing a print. They chose to incorporate an all-in-one solution using the Duet 3 controller board and its CAN bus expansion board respectively for normal operations of the 3D printer and tensile testing operations. The authors express the ease of modularity with the CAN bus for the expansion of further tasks. Fig. \ref{fig:tensile3d} shows a 3D render of the integrated tensile testing and 3D printer. 
\subsection{Case Study 3: Mobile Large Scale Concrete 3D Printer Design}
\begin{figure}[!ht]
\vspace*{-10pt}
	\centering	\includegraphics[trim=5cm 8cm 5cm 7.9cm, width=2.5in, clip]{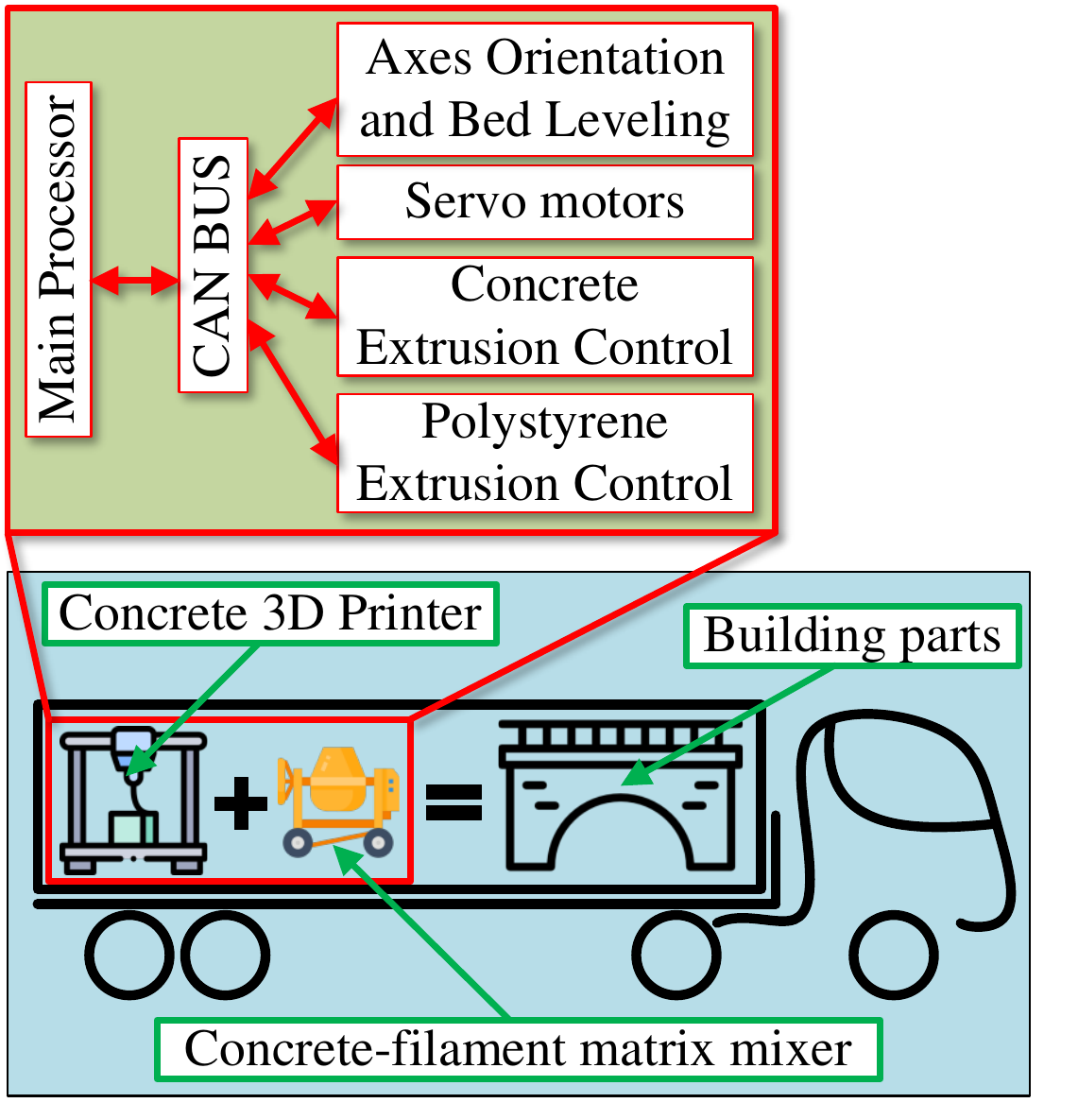}
	\caption{Mobile Cement 3D Printer Concept \cite{QQ1}.}	
	\label{fig:cement3d}
\end{figure}

Recently, the researchers in \cite{TT1} list 42 architectural works utilizing concrete 3D printing, and provided an exhaustive survey of the advantages and limitations of this manufacturing technique. For example, Habitat for Humanity, a prominent non-profit construction organization, is venturing into the arena of 3D-printed houses \cite{RR1}. The number of buildings constructed by these companies points to the rapid adoption and commercialization of AM within the construction industry.
 
A group of engineering students at the University of Surrey proposed a design for a portable concrete 3D printer that is housed in a semi-truck trailer \cite{PP1}. Their proposed design is shown in fig. \ref{fig:cement3d}. This is an interesting idea, as concrete 3D printing is akin to conventional 3D printing, which requires a more delicate operation. Essentially, the locations of the print head, bed leveling, and cement composite status have to be considered. The significant difference between a conventional 3D printer and a cement 3D printer is the size and print material. A CAN bus is capable of transmitting over huge distances without compromising speed or data accuracy and is capable of maintaining and monitoring the current status of the printer due to its masterless nature.    
\subsection{Case Study 4: Fiber Orientation Control via Rotational Print Heads in 3D Printing}
%
M.D. Roberts designed and implemented a prototype rotational print head to control fiber orientation in 3D printing \cite{UU1}. This process is called Rotational Fused Filament Fabrication (RFFF), in which fiber orientation is controlled by inducing rotational shear forces into the flow of material through the nozzle. Fiber orientation is controlled through changes in the shear rate, allowing for the printing of structures with targeted local properties.     

The main controllers used in this prototype are a combination of the Duet2 WiFi and the Hyrel 3D Hydra 16A control board. The Duet2 board is used for the extruder control and rotational fiber orientation control, while the Hydra 16A board is used for controlling the XYZ axis of the 3D printer system. By inspecting the provided wiring diagram, data monitoring and syncing are performed over the CAN bus where the status of rotation, temperature, and filament are shared between the controllers.   




\section{DISCUSSION AND FUTURE RESEARCH DIRECTION}
The integration of AM with CAN has the potential to be the next step for the normalization of its inter-hardware communication protocol. Users of AM span from enthusiasts at home to professionals in laboratories around the world. Upgrades are often sought-after for their equipment to expand their options of what they can create. By using an AM system that has CAN as its main choice of central communication, the users can take advantage of the modularity and ease of data monitoring. Similar to how the automotive industry provides ECUs with a lot of low-level functionality on the CAN bus, the robustness and ease of use should be utilized in the AM market as well. 

Even though CAN provides live, high-speed communication between connected hardware, it still has disadvantages associated with it. The first drawback is the CAN bus wire length to signaling rate limitation, with a recommended maximum of 30 nodes. More nodes can be added, extending wire lengths and causing trade-off considerations to signal rates \cite{JJ1}. Another issue is that any physical access to the CAN bus provides access to its multi-destination input and output. Thus, the CAN bus is also exploitable due to not having pre-built provisions for authentication and confidentiality, which makes it vulnerable\cite{KK1, LL1, WW1, XX1}. The march forward into future generations of the manufacturing industry and the IoT are prone to security breaches which have huge consequences. Any future applications with CAN have to check and cover all vulnerabilities to safeguard lives, property, and the environment from malicious entities.

\begin{IEEEbiography}{Jun-Cheng Chin}{\,}is currently working toward the M.S.
degree in electrical engineering with the University of
Tennessee, Knoxville, TN 37996, USA. Contact him at
jchin2@vols.utk.edu.
\end{IEEEbiography}

\begin{IEEEbiography}{Himanshu Thapliyal}{\,} is an Associate Professor with the
Department of Electrical Engineering {\&} Computer Science,
University of Tennessee, Knoxville, TN 37996, USA. Contact
him at hthapliyal@utk.edu.
\end{IEEEbiography}

\begin{IEEEbiography}{Tyler Cultice} {\,} is currently working toward the Ph.D.
degree in electrical and computer engineering with the University
of Tennessee, Knoxville, TN 37996, USA. Contact him at
tcultice@vols.utk.edu.
\end{IEEEbiography}

\end{document}